%% file: main.tex
\documentclass[journal]{IEEEtran}

\usepackage{cite}
\usepackage{amsmath,amssymb,amsfonts}
\usepackage{algorithm}
\usepackage{algorithmic}
\usepackage{makecell}
\usepackage{threeparttable}
\usepackage{graphicx}
\usepackage{textcomp}
\usepackage{xcolor}
\usepackage{multirow}
\usepackage{dsfont}
\usepackage{url}

\ifCLASSINFOpdf
\else
\fi
\begin{document}
\title{Seeking Common Ground While Reserving Differences: Multiple Anatomy Collaborative Framework for Undersampled MRI Reconstruction}

\author{Jiangpeng Yan, Chenghui Yu, Hanbo Chen, Zhe Xu, Junzhou Huang, \IEEEmembership{Member, IEEE}, Xiu Li, \IEEEmembership{Member, IEEE}, and Jianhua Yao, \IEEEmembership{Senior Member, IEEE}
\thanks{Manuscript received XXX; revised XXX; accepted XXX. Date of publication XXX; date of current version XXX. This research was partly supported by the XXX.}
\thanks{J. Yan, H. Yu and H. Chen have equal contributions to the work. X. Li and J. Yao are corresponding authors.}
\thanks{J. Yan, H. Yu and X. Li are with Tsinghua Shenzhen International Graduate School, Tsinghua University, Shenzhen 518055, China. (E-mail: yanjp17@mails.tsinghua.edu.cn; li.xiu@sz.tsinghua.edu.cn).}
\thanks{H. Chen and J. Yao are with 
AI healthcare in Tencent AI Lab, Shenzhen 518055, China. (E-mail: hanbochen@tencent.com; jianhuayao@tencent.com).}
\thanks{Z. Xu is with Department of Biomedical Engineering, The Chinese University of Hong Kong, Shatin, NT, Hong Kong, China. (E-mail: jackxz@link.cuhk.edu.hk).}
\thanks{J. Huang is with the Department of Computer Science and Engineering,
University of Texas at Arlington. (E-mail: jzhuang@uta.edu).}
}


\markboth{Journal of \LaTeX\ Class Files,~Vol.~14, No.~8, August~2015}%
{Shell \MakeLowercase{\textit{et al.}}: Bare Demo of IEEEtran.cls for IEEE Journals}



\maketitle

\begin{abstract}
Recently, deep neural networks have greatly advanced undersampled Magnetic Resonance Image (MRI) reconstruction, wherein most studies follow the one-anatomy-one-network fashion, i.e., each expert network is trained and evaluated for a specific anatomy.
Apart from inefficiency in training multiple independent models, such convention ignores the shared de-aliasing knowledge across various anatomies which can benefit each other. To explore the shared knowledge, one naive way is to combine all the data from various anatomies to train an all-round network. Unfortunately, despite the existence of the shared de-aliasing knowledge, we reveal that the exclusive knowledge across different anatomies can deteriorate specific reconstruction targets, yielding overall performance degradation. Observing this, in this study, we present a novel deep MRI reconstruction framework with both anatomy-shared and anatomy-specific parameterized learners, aiming to ``seek common ground while reserving differences'' across different anatomies.
Particularly, the primary anatomy-shared learners are exposed to different anatomies to model flourishing shared knowledge, while the efficient anatomy-specific learners are trained with their target anatomy for exclusive knowledge. Four different implementations of anatomy-specific learners are presented and explored on the top of our framework in two MRI reconstruction networks. Comprehensive experiments on brain, knee and cardiac MRI datasets demonstrate that three of these learners are able to enhance reconstruction performance via multiple anatomy collaborative learning.
\end{abstract}


% \fi
\IEEEpeerreviewmaketitle

\section{Introduction}
\label{sec:intro}

\begin{figure*}
\centering
\includegraphics[width=\textwidth]{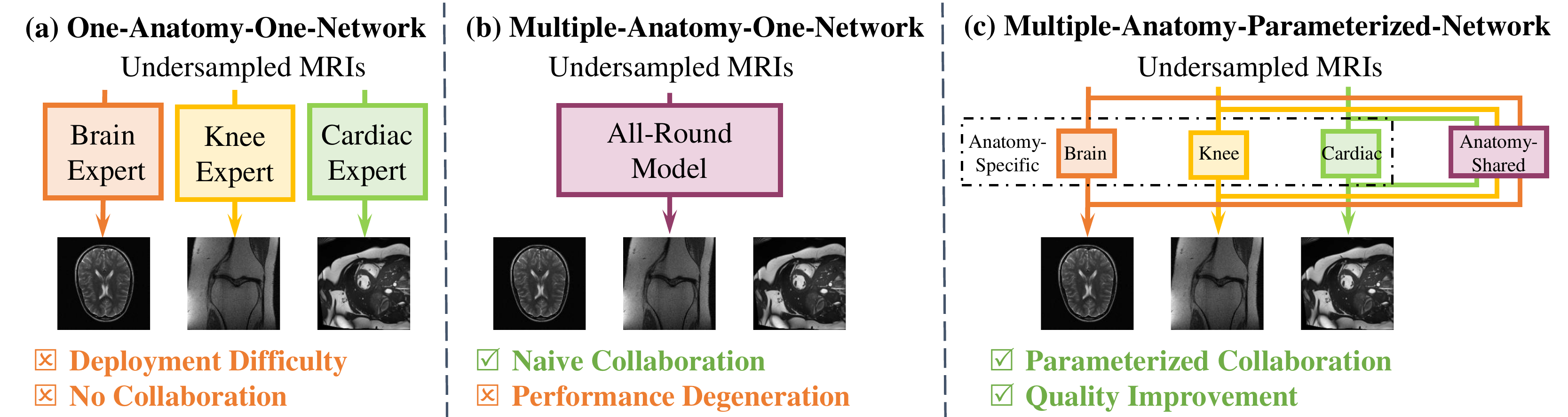}
\caption{Comparison of (a) One-Anatomy-One-Network, (b) Multiple-Anatomy-One-Network, and (c) Our Proposed Multiple-Anatomy-Parameterized-Network paradigm.}
\label{fig_intro}
\end{figure*}

\IEEEPARstart{M}{agnetic} Resonance Images (MRIs), which can provide detailed anatomical information without invasion, are widely used for various image-guided therapies. As a typical compressed-sensing \cite{donoho2006compressed} based inverse problem, reconstructing high-quality MRIs from undersampled k-space signals plays an important role in accelerating the imaging process \cite{liang2009accelerating,huang2011efficient}. Recently, deep convolutional neural networks (CNNs) have proven soaring successes in efficiently reconstructing high-quality de-aliased anatomy MRIs \cite{liang2020deep,knoll2020deep} with their stronger ability of automatically learning high-level anatomical representation, wherein most efforts \cite{wang2016accelerating,schlemper2017deep,qin2018convolutional,hammernik2018learning,aggarwal2018modl,lee2018deep,sun2018compressed,zheng2019cascaded,yan2020neural,sriram2020end,jun2021joint} follow the one-anatomy-one-network (OAON) fashion, i.e., each expert network is trained and evaluated for specific anatomy, as shown in Fig. \ref{fig_intro} (a). However, training multiple independent networks for different anatomies greatly increases the deployment difficulty since tens of anatomies are examined with MRIs in clinical practice. Moreover, apart from inefficiency in deployment, such convention ignores the shared de-aliasing knowledge across various anatomies, which can benefit each other for reconstruction. 

Therefore, it is natural to ask: \textbf{whether a deep network can simultaneously learn from mixed anatomy data for all-round MRI reconstruction?} Intuitively, one naive solution is to adopt a multiple-anatomy-one-network (MAON) fashion that directly introduces all the data from various anatomies to train an all-round network, as shown in Fig. \ref{fig_intro} (b). Unfortunately, despite the existence of the shared de-aliasing knowledge and better generalization ability, we reveal that the exclusive knowledge across different anatomies (e.g., unique intensity distributions and structure/location priors) can mislead the network and deteriorate specific reconstruction targets, yielding overall decreased reconstruction quality. 

Most relevantly, Liu et al. \cite{liu2021universal} introduced a universal undersampled MRI reconstruction network trained by distilling knowledge from pre-trained expert models into one universal network and the anatomy-specific instance normalization was proposed to adapt the various intensity distributions for multiple anatomies. Given the knowledge distilling as a  complicated process where pre-trained expert models are still required, here we lean to disentangle knowledge via different parameterized learners. In terms of the anatomy-specific normalization, we make a step further than \cite{liu2021universal} to use the parameterized learners for capturing anatomy-specific information with a novel unified framework, where anatomy-specific normalization can be regarded as one of the possible implementations.

In this study, we propose to seek common ground while reserving differences across different anatomies by redesigning a novel deep MRI reconstruction framework constructed with anatomy-shared and anatomy-specific parameterized learners, named as multiple-anatomy-parameterized-network (MAPN) framework shown in Fig. \ref{fig_intro} (c). Particularly, such design echos the multi-domain visual classification learning \cite{rebuffi17,rebuffi18} and allows the MRI reconstruction of each anatomy to enjoy both (i) the flourishing shared de-aliasing knowledge by training the primary anatomy-shared learners with mixed anatomies, and (ii) the exclusive knowledge reserved for it in the tailored anatomy-specific learners. In fact, the OAON and MAON fashion are the special cases of our MAPN where there exist either anatomy-specific or anatomy-shared learners. Under our proposed unified framework, we explore four types of anatomy-specific learners on top of our framework, including anatomy-specific normalization \cite{liu2021universal}, channel attention, series learners and paralleled learners, and integrated them into two different MRI reconstruction networks. Comprehensive experiments on brain, knee and cardiac MRI datasets demonstrate the effectiveness of our new ``all-round" network with various insightful findings. The contributions of this work can be summarized as follows:

\begin{itemize}
    \item We unify current one-anatomy-one-network and multiple-anatomy-one-network solutions under a novel multiple-anatomy-parameterized-network framework where we can both seek common ground and reserve differences across anatomies.
    \item We implement four possible designs of the compact anatomy-specific parameterized learners as the early attempts under the proposed framework and three of them show the ability to enhance multiple anatomy collaborative learning with two different reconstruction networks, providing insightful findings on how to better capture anatomy-specific information with our framework.
    \item Extensive experiments are performed on brain, knee and cardiac MRI datasets demonstrating the effectiveness of our strategy for improving different anatomy reconstruction quality with different reconstruction networks.
\end{itemize}

\section{Related Work}

\subsection{Deep Learning Based Undersampled MRI Reconstruction}
Since Wang et al.\cite{wang2016accelerating} firstly introduced deep networks for undersampled MRI reconstruction, researchers have developed various approaches which can be roughly categorized into the data-driven \cite{wang2016accelerating,fastMRI,zhu2018image,yang2017dagan,lee2018deep}, the model-driven \cite{hammernik2018learning,aggarwal2018modl,akccakaya2019scan,schlemper2017deep,qin2018convolutional,jun2021joint,zhang2021dual}. Among these works, DCCNN \cite{schlemper2017deep} has gained much attention with a series of variants \cite{sun2018compressed,zheng2019cascaded,yan2020neural,sriram2020end,jun2021joint} where researchers explored how to design neural networks for better representation learning. However, these methods all followed the OAON fashion as mentioned above, and there only existed a few previous works discussing the anatomy generalization ability in MRI reconstruction. Yan et al. \cite{yan2020neural} used neural architecture search technology to search different network architectures for knee and brain, proving that different anatomies prefer different architectures. Transfer learning in MRI reconstruction is discussed in \cite{ARSHAD202196, LV2021104504,dar2020transfer}, where researchers pre-trained a network on a large pool of data and then transferred the learned weights for one anatomy reconstruction by fine-tuning, while our aim is to learn and reconstruct multiple anatomies simultaneously in one network. Guo et al. \cite{guo2021multi} introduced the distribution shifting problem of MRIs from different institutions and proposed a federated learning-based solution but only for brain MRI reconstruction tasks. Few attempts were made to explore how different anatomies can be reconstructed by one network. 

As mentioned above, Liu et al. \cite{liu2021universal} made the first step to introduce a universal undersampled MRI reconstruction network with knowledge distilling and the anatomy-specific instance normalization strategy. In our framework, we lean to disentangle knowledge via different parameterized learners without the need of difficult knowledge distilling process. Since the idea of anatomy-specific instance normalization can be regarded as a setting of the anatomy-specific learners, we re-implemented \cite{liu2021universal} as the anatomy-specific normalization settings without knowledge distilling to make a fair comparison with other settings in our experiments. 

\subsection{Multi-domain Visual Classification Learning}

Our method is closely related to recent works in multi-domain visual classification learning \cite{bilen2017universal,rebuffi17,rebuffi18,mallya2018piggyback}, where researchers aim to train a single network to perform image classification tasks across different domains (animals, flowers, digits, ...). Bilen and Vedaldi \cite{bilen2017universal} used parameters in batch and instance normalization layers to model different domains for task-specific classification. Rebuffi et al. extended \cite{bilen2017universal} in \cite{rebuffi17,rebuffi18} to parameterize the standard residual network architecture into a network family to capture different domains. Mallya et al. \cite{mallya2018piggyback} proposed to adapt a single network to different classification tasks by masking weights of different convolution kernels. 

Distinct from the above works, our aim is to make MRI reconstruction model jointly and better learn from multiple anatomies data and we make the early attempt to introduce the idea in \cite{rebuffi17,rebuffi18} for building an all-round MRI reconstruction model with a parameterized network where the compact anatomy-specific learners focus on exclusive anatomy prior while the major anatomy-shared learners enjoy common de-aliasing knowledge. Our experiments demonstrate that such setting can help different anatomies collaborate with each other in the simultaneous learning process.

\section{Methodology}

\begin{figure*}
\centering
\includegraphics[width=\textwidth]{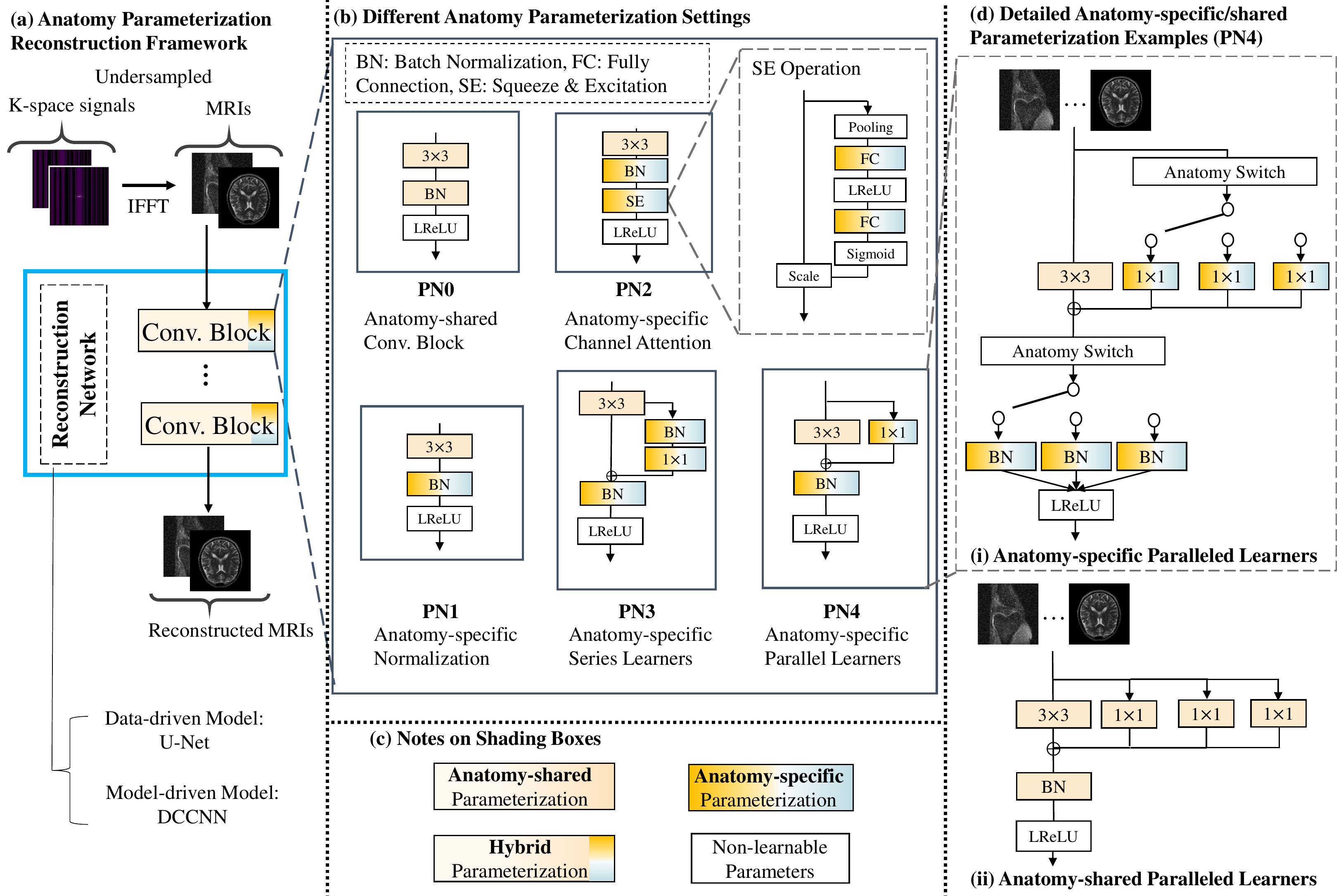}
\caption{Illustration of (a) Anatomy parameterization reconstruction framework for multiple anatomy collaborative learning. Our framework can be generalized to both data-driven and model-driven models. (b) Details of four anatomy parameterization settings. (c) Notes on shading boxes. (d) Anatomy-specific parameterization example with paralleled learners setting (d-i) and its anatomy-shared variant (d-ii) for the ablation study.}
\label{fig_framework}
\end{figure*}

Without loss of generalization ability, we study multiple anatomy undersampled MRI reconstruction problems under single-coil reconstruction scenario following \cite{sun2018compressed,zheng2019cascaded,liu2021universal}. We firstly introduce the problem formulation and two baseline solutions: the data-driven U-Net \cite{ronneberger2015u} based reconstruction model \cite{fastMRI,yang2017dagan,lee2018deep} and the model-driven deep cascaded network (DCCNN) based reconstruction model \cite{schlemper2017deep,sun2018compressed,zheng2019cascaded,yan2020neural,liu2021universal}. Then, we illustrate the details of the proposed MAPN solution with four anatomy-specific learner designs. The overall framework is shown in Fig. \ref{fig_framework}.

\subsection{Problem Formulation and Baseline Solutions}

The aim of MRI reconstruction is to recover the fully-sampled MR image $x$ from the under-sampled k-space signals $s$, which can be formulated as:
\begin{equation}
\label{equ:1}
s = F_{u} x,
\end{equation}
where $F_{u}$ is the under-sampling Fourier encoding matrix and $x$ can be obtained by solving the following inverse problem:
\begin{equation}
\label{equ:2}
\min_x \lVert s - F_{u} x\rVert_2^2 + \mathcal{R}(x),
\end{equation}
where $\mathcal{R}(x)$ is the regularization term. By integrating CNN-derived constraints into Eq.~\ref{equ:2}, this problem can be formulated as:
\begin{equation}
\label{equ:3}
\min_x  \lVert s - F_{u} x\rVert_2^{2} + \lambda \lVert x - \mathcal{N} \left( x , \omega )\right \rVert^2,
\end{equation}
where $\mathcal{N}$ represents the deep CNN network with learn-able weights $\omega$.

For data-driven solutions, researchers usually solve Eq.~\ref{equ:3} by directly learning a mapping between between paired under-sampled MRI $x_u$ and fully-sampled MRI $x$. Typically, U-Net \cite{ronneberger2015u} is adapted in \cite{fastMRI,yang2017dagan,lee2018deep} as $\mathcal{N}$ and we have: 
\begin{equation}
\label{equ:u}
y = \mathcal{N}( x_u , \omega).
\end{equation}
Then, $\mathcal{N}$ is trained to reconstruction $x$ from $x_u$ by minimizing the objective function $\mathcal{L}$ as: 
\begin{equation}
\mathop{{\min}}\limits_{{\omega}} \mathcal{L} (\mathcal{N}(x_{u}, \omega)-x),
\end{equation}
where $\mathcal{L}$ is the pixel-wise loss, e.g., L1 or L2 loss.

For model-driven solutions, researchers tend to regard $\lVert s - F_{u} x\rVert_2^2$ as the data consistence (DC) term between the image and k-space data, then unroll Eq.~\ref{equ:3} with the alternating minimization steps \cite{schlemper2017deep,sun2018compressed,zheng2019cascaded,yan2020neural,liu2021universal}. Schlemper et al. \cite{schlemper2017deep} firstly proposed to use the data consistency (DC) modules and re-formulated Eq.~\ref{equ:3} by cascading $n$ sub-networks as the following two steps:
\begin{equation}
\label{equ:k}
y^{(n)} = \mathcal{N}^{(n)}( x^{(n)} , \omega^{(n)} ),
\end{equation}
\begin{equation}
\label{equ:m}
x^{(n+1)} = \arg\min_x  \lambda \lVert s_u - F_{u} x\rVert_2^{2} + \lVert x - y^{(n)}\rVert^2,
\end{equation}
where $x^{(0)}=x_u$. The problem of Eq. \ref{equ:m} is given with a closed-form solution with the DC layer, which can be regarded as a plug-in module in the deep network to align current reconstructed k-space of $y^{(n)}$ with the ground truth sampled signals in $s$, given $\lambda \rightarrow \infty$. The problem of Eq. \ref{equ:k} is similar to Eq. \ref{equ:u}. A super network $\mathcal{N}$ is built with $n$-fold cascaded sub-CNNs $\mathcal{N}^{(n)}$ and DC modules, and trained by minimizing the objective function $\mathcal{L}$ to solve Eq. \ref{equ:k}. 

\begin{figure*}
\centering
\includegraphics[width=\textwidth]{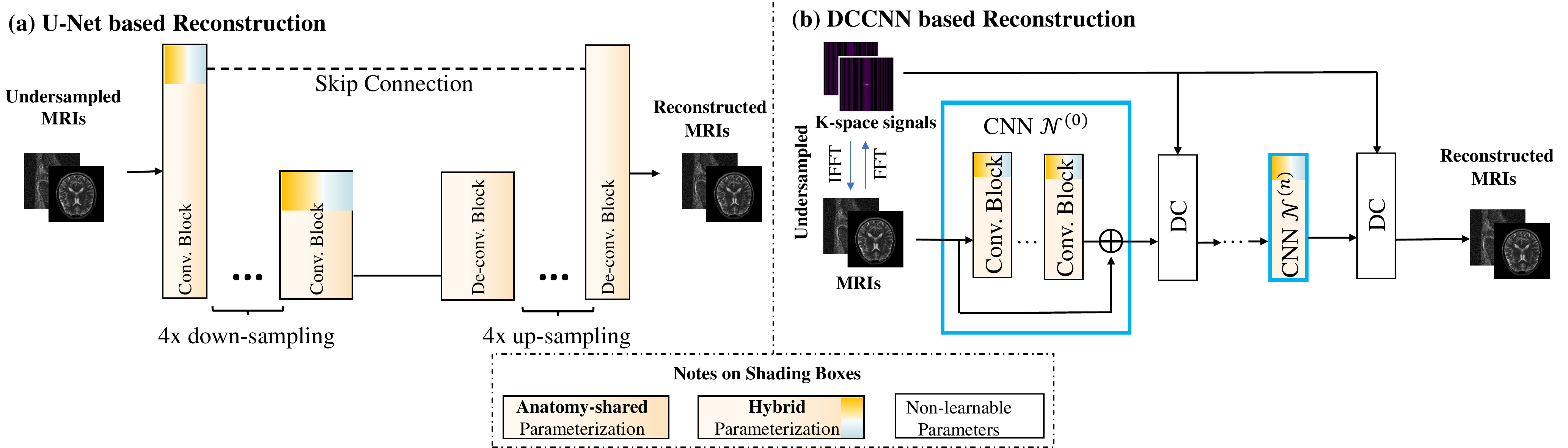}
\caption{Illustration of Anatomy Parameterization based U-Net \cite{fastMRI} (a) and DCCNN \cite{schlemper2017deep} (b) for multiple anatomy collaborative learning.}
\label{fig_unet}
\end{figure*}

For better understanding, we visualize U-Net based reconstruction network in Fig.\ref{fig_unet} (a), where we follow the setting in \cite{fastMRI} with 4 down/up-sampling operations, and DCCNN based reconstruction network in Fig.\ref{fig_unet} (b) with our proposed anatomy parameterization strategy. At a higher level, we can regard the U-Net based reconstruction solution as a sub-problem of the DCCNN based one, and U-Net \cite{ronneberger2015u} can be directly used as the sub-CNNs $\mathcal{N}^{(n)}$ in DCCNN. Here, we independently treat these two solutions to demonstrate the generalization ability of our anatomy parameterization framework. What's more, in original DCCNN \cite{schlemper2017deep}, the researchers only use plain convolutional layers to build their network. Comparing to \cite{schlemper2017deep}, the U-Net includes 4 down-sampling blocks with convolution layers and 4 up-sampling blocks with de-convolution layers. For simplicity and to keep consistent with DCCNN, we replace all the convolution layers in U-Net with anatomy-shared/specific learners and leave the de-convolution layers as anatomy-shared learners.

%Since DCCNN only uses plain convolution layers for feature extraction, we also explore if our framework can be adapted to other networks. We evaluate the possibility of integrating our framework with the Decode-Encoder reconstruction networks, i.e. U-Net\cite{ronneberger2015u} based solutions \cite{yang2017dagan,lee2018deep,fastMRI}, where contains down-sampling and deconvolution operations. In these methods, researchers use U-Net\cite{ronneberger2015u} as $\mathcal{N}$ to directly learn a mapping between $x_u$ and $x$ under supervision of $\mathcal{L}$ with:

Notably, most existed works follow the OAON setting, where one independent $\mathcal{N}$ is trained from scratch and only tested for one target reconstruction anatomy. In this work, we benchmark above baseline solutions with multiple anatomy reconstruction task.

%Notably, Batch Normalization (BN) layers \cite{ioffe2015batch} are adopted in the original DCCNN \cite{schlemper2017deep} and its follow-up works \cite{sun2018compressed,zheng2019cascaded,qin2018convolutional,yan2020neural} while there also exist some works \cite{fastMRI,zhang2021dual,liu2021universal} used Instance Normalization (IN) \cite{ulyanov2016instance} instead of BN. As comparing different normalization settings is beyond our scope in this work, we use BN in all the experiments. Our focus is to study the multiple anatomy MRI reconstruction problem with such a framework.

\subsection{Multiple Anatomy Parameterized Framework}
Revisiting Eq. \ref{equ:u} and Eq. \ref{equ:k} for multiple-anatomy reconstruction scenario, it is a natural idea to disentangle $\omega$ with the whole network $\mathcal{N}$ for the shared and exclusive knowledge via rewriting Eq. \ref{equ:u} and Eq. \ref{equ:k} as:
\begin{equation}
\label{equ:f}
y = \mathcal{N}( x_u, \omega_s, \omega_{e^i}),
\end{equation}
where $\omega_s$ is trained with all anatomies to learn shared de-aliasing knowledge and $\omega_{e^i}$ is from a parametric set $\Omega_e=\{\omega_{e^1},\omega_{e^2},...,\omega_{e^n}\}$ in which each element is corresponding to a specific anatomy. Examples of detailed implementations of $\omega_s$ and $\Omega_e$ are visualized in Fig. \ref{fig_framework} (d) for better understanding, termed as anatomy-shared/specific parameterization. With Eq. \ref{equ:f} as the basis of our MAPN framework, we then need to answer: \textbf{how to properly design $\Omega_e$?}

The design of $\Omega_e$ needs to consider the following two criteria: (1) $\omega_{e^i}$ should be more efficient compared to $\omega_s$, otherwise it brings in more learning parameters; (2) the $\omega_{e^i}$ should be easy to plug into and out from the network with $\omega_s$ such that they can be easily switched for different organs and enjoy the shared knowledge. 

With the above two criteria, when we further investigate the convolution blocks shown in Fig. \ref{fig_framework}
(b) PN0, we can find that an intuitive setting is to use the $3\times3$ convolutional layers as $\omega_{s}$, while the common Batch Normalization (BN) \cite{ioffe2015batch} layers can be regarded as an anatomy-specific setting of $\omega_{e^i}$ shown in Fig. \ref{fig_framework}
(b) PN1 noted by Anatomy-specific Normalization strategy. It is obvious that MRIs of different anatomies have significant feature distribution shifting in terms of means and variances. Given the input feature map as $m$, the normalization layers utilized to calibrate the feature map can be formulated as:
\begin{equation}
n = \gamma \cdot \frac{ m - \mu(m)}{\sigma(m)} + \beta,
\end{equation}
where scale weight $\gamma$, shift bias $\beta$ in BN layers. To make BN layers better adapt to one specific anatomy with its unique intensity distribution, one independent set of normalization layers is assigned with unique learning parameters $[\gamma, \beta]$ by an anatomy switch operation in the network for each anatomy, as indicting in Fig. \ref{fig_framework} (d-i). It is reminded that the anatomy-specific instance normalization used in \cite{liu2021universal} is similar to PN1 setting with $[\mu(m), \sigma(m)]$ calculated differently and a complicated knowledge distillation step. With a fair comparison protocol comparing to following other settings, we think PN1 is our enhanced re-implemented version of \cite{liu2021universal}, given codes/models of \cite{liu2021universal} are not open source. 

With Anatomy-specific Normalization as a good starting point, another direction to design $\Omega_{e}$  is to plug some additive compact tailored anatomy-specific structure into the convolution block for exclusive knowledge learning. We propose to use the channel attention mechanism, which was introduced in Squeeze-and-Excited (SE) Network \cite{hu2018squeeze}, to model anatomy knowledge by re-calibrating channel-wise feature responses with fully-connected layers shown in Fig. \ref{fig_framework} (b) PN2 noted by Anatomy-specific Channel Attention strategy. 

So far, the previous two solutions do not change the major graphic structure of the network. In order to leverage features from different levels and inspired by successes in multi-domain natural image classification \cite{rebuffi17,rebuffi18}, we use $1\times1$ convolutional layers as residual connections to adjust features from different levels and then obtain the setting in Fig. \ref{fig_framework} (b) PN3 and PN4, namely Anatomy-specific Series and Parallel Learners, correspondingly, according to where the learners locate. The $1\times1$ convolutional operations are chosen due to parameter efficiency. 

The above four settings are explored as possible implementations under our proposed framework to prove its effectiveness and how to design the optimal anatomy-specific learners remains an open problem. We explore these implementations to shed the light in a new direction for multiple anatomy collaborative reconstruction. 

%The series and parallel learners echo previous works \cite{rebuffi17,rebuffi18} where researchers used such structures for multi-domain natural image classification, while we introduce such settings on MRI reconstruction for the first time. 

%In the above four settings, the PN1 acts as the basic setting upon which PN2/PN3/PN4 are then built. PN2/PN3/PN4 can be further integrated together flexibly but with heavier parameter burden. In this study, we use the above four settings independently to demonstrate the potential of our proposed framework.

\section{Experiments}
\label{sec:exp}
\subsection{Implementation Details}
\subsubsection{Environment} All the experiments were carried out on an NVIDIA GeForce RTX 3090 GPU with the Pytorch.
\subsubsection{Dataset Preprocessing and Evaluation Metrics} We conducted our experiments on three public MRI datasets of different anatomies with Institutional Review Board approval: \textbf{Knee}: we randomly sampled 800 knee MR slices from 25 subjects in the fastMRI dataset \cite{fastMRI} provided by New York University School of Medicine. \textbf{Brain}: we randomly sampled 800 brain MR slices from 50 subjects in the fastMRI dataset \cite{fastMRI}. \textbf{Cardiac:} we randomly sampled 800 cardiac MR slices from 90 cardiac subjects in the ACDC dataset \cite{bernard2018deep} obtained from the University Hospital of Dijon. Noteworthily, we removed some slices where no biological structure exists to make three anatomies have equal slices for fairness. It is reminded that the above three datasets are acquired with different 
protocols, and we follow the setting in \cite{liu2021universal} to treat them as the same single-coil slice-level MR image for reconstruction.
Following \cite{sun2018compressed,zheng2019cascaded,liu2021universal}, we adopted these data pre-processing steps: the MRIs were cropped into $320 \times 320$, then normalized in magnitude, and clipped to [-6, 6] with zero phases to simulate 2-channel complex-valued images. 
%The $4\times$ under-sampled 1D Cartesian masks were adopted on Fast-Fourier-Transformed (FFT) MRIs to produce corresponding under-sampled k-space signals, i.e., the input under-sampled MRIs via IFFT. 
%Particularly, the central $8\%$ k-space lines were full-sampled as autocalibrated
%signals (ACS) with the outer $(25-8=17\%)$ k-space under-sampled uniformly. 
All the models were trained with under-sampled 1D Cartesian masks and evaluated with  $4\times$ and $6\times$ under-sampled reconstruction tasks.
We report PSNR and SSIM \cite{hore2010image} between reconstructed and target fully-sampling MRIs as the quantitative metrics. The final results are calculated on a 5-fold cross-validation setting by dividing the above three datasets according to subjects with a ratio of 4:1 for the training and validation.
\subsubsection{Network Setting} To implement DCCNN reconstruction network, we used 5 stacked convolutional blocks to build each sub-CNN and cascaded the sub-CNNs and DC modules for 5 times to build the baseline DCCNN. Each block includes cascading Conv.-BN-LeakyReLU operations, as shown in Fig. \ref{fig_framework} (b) PN0, with 64 channels. We followed the setting in \cite{fastMRI} to build a U-Net based reconstruction network but modified the input and output as 2 channels to keep consistent with the DCCNN. To perform anatomy parameterization settings, we replaced all the convolutional operations, i.e. PN0, in U-Net/DCCNN by PN1-4, as shown in Fig. \ref{fig_framework} (b).

\subsubsection{Training Strategies} To implement the OAON baselines, we trained U-Net/DCCNN based reconstruction models independently on three datasets for 50 epochs. 
To implement the multiple-anatomy-related baseline, we fed a batch of data from the same anatomy in a round-robin fashion in each iteration for $3\times50=150$ epochs as we used 3 anatomies in this study.
For the MAPN methods, the anatomy-specific learners were optimized only once in each iteration for their target anatomy, while the anatomy-shared $3\times3$ convolution layers were optimized in every mini-batch across anatomies. We also adopted a warm-up strategy for anatomy parameterized learners to dig out exclusive knowledge: we loaded and fixed the weight of $3\times3$ convolution layers obtained from the MAON baseline for the first 30 epochs to better initialize the additive anatomy-specific modules, then the whole network was fully optimized for the next 120 epochs. The L1 loss was adopted for supervision and the Adam optimizer with the same learning rate of 0.01 was adopted for all the methods.

\subsection{Experiments of U-Net based Reconstruction} 

Here, we report the reconstruction performance of the anatomy parameterization based U-Net as shown in Fig. \ref{fig_unet} (a). 

\subsubsection{Quantitative Results} We summarized the quantitative results in Table. \ref{tab_3}, which can be divided into three parts under $4\times$ and $6\times$ under-sampling ratio: (i) OAON setting where knee/brain/cardiac expert models are trained independently, (ii) MAON setting where all-round models are trained on the mixture of three datasets, (iii) four different MAPN settings. Though the overall performance with $6\times$ under-sampling ratio is lower than that with $4\times$ ratio, findings are quite consistent between them, suggesting the robustness of proposed method. Intuitively, we can find that MRI reconstruction results can not benefit from direct multiple anatomy data mixed training. Expert for brain reconstruction with $4\times$ ratio, the knee/brain/cardiac expert models all outperform the MAON model although with unsatisfied anatomy generalization ability. Our proposed MAPN solutions can learn from different anatomy to better reconstruct MRIs compared to the naive MAON solution. By integrating anatomy-specific batch normalization (PN1), we can find that the network can enjoy the shared knowledge learned in $3\times3$ convolutional layers to outperform the MAON baseline in all three tasks but can not provide better results than OAON experts in terms of the cardiac reconstructions. After integrating anatomy-specific learners with PN2-4 settings, the MAPN network can reconstruct all three anatomies better than both OAON expert models and MAON baselines from collaborative learning. Notably, the PN2 setting achieves comparable results with PN4 settings in U-Net based reconstruction.

\subsubsection{Qualitative Results} 
\begin{figure*}
\centering
\includegraphics[width=\textwidth]{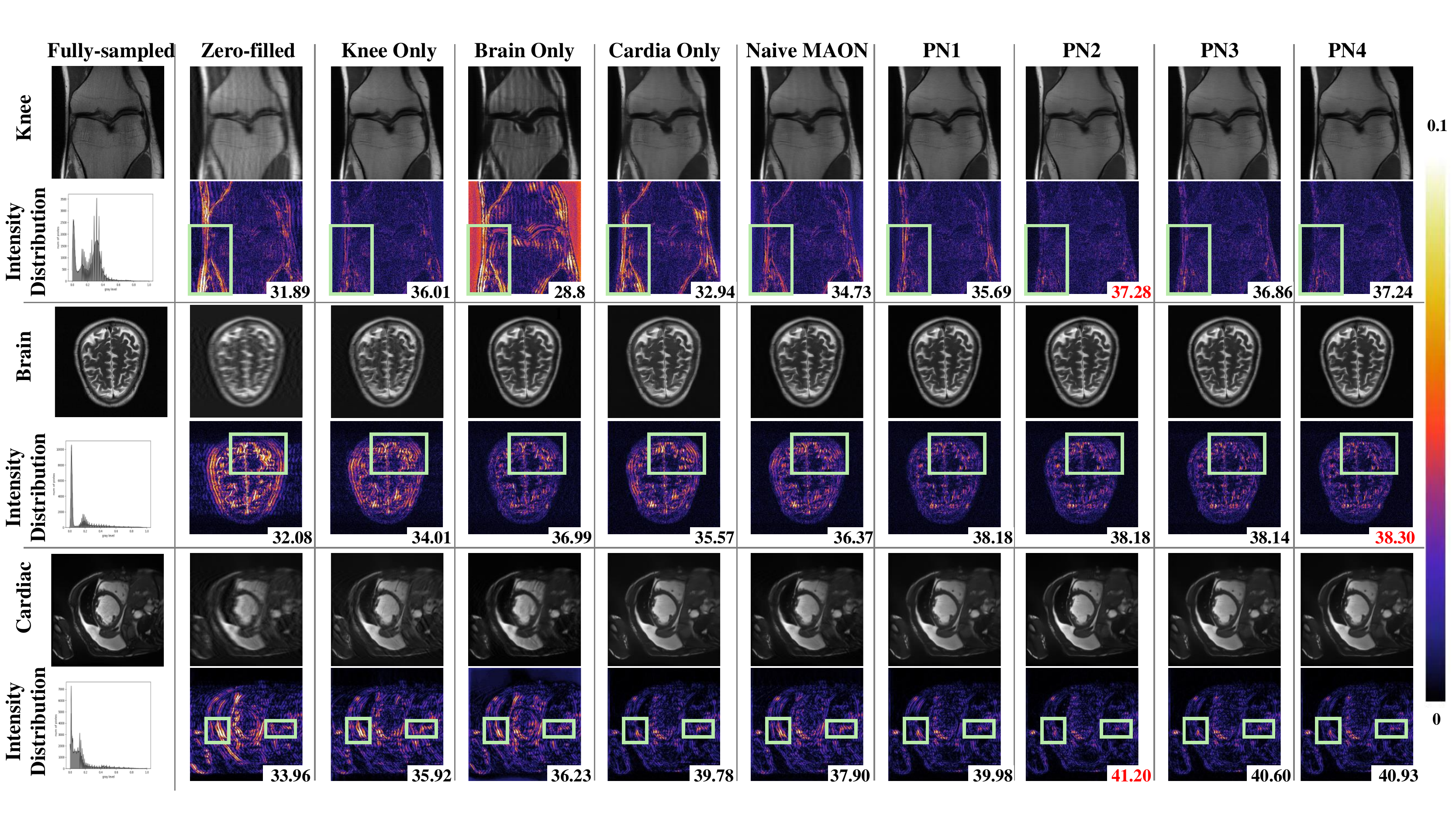}
\caption{Qualitative comparison of different anatomy parameterization U-Net reconstruction settings with $4\times$ under-sampling ratio. Original MRIs are normalized to [0, 1] for visualization and their intensity distributions are shown in the first column. From the second column, reconstructed image (top) and its absolute error map (bottom) are shown. PSNR (dB) is listed at the right bottom of each error map with the highest one highlighted in red. Values $>0.1$ in error maps are clipped to 0.1 for better visual comparison. Green boxes address where significant performance difference can be observed.}
\label{fig_comparison_u}
\end{figure*}

We also visualize reconstruction performance of all settings in U-Net with $4\times$ ratio in Fig. \ref{fig_comparison_u}. Similar to the qualitative results, we can find that networks with our parameterized learners can reconstruct images with clearer details and reduced error maps.

\begin{table*}
\caption{Quantitative reconstruction results of U-Net based reconstruction on 5-fold cross validation of different anatomy parameterization strategies with the metric of PSNR(dB) and SSIM(\%). `Und. Rat.' represents Under-sampling Ratio. `K',`B', and `C' are short for `Knee', `Brain' and `Cardiac' at the 3rd column for compactness. `OAON' is short for `One-Anatomy-One-Network'. `MAON' is short for `Multiple-Anatomy-One-Network'. `MAPN' is short for `Multiple-Anatomy-Parameterized-Network'. The best results are addressed in bold.}
\label{tab_3}
\centering
\scalebox{0.9}{
\begin{tabular}{c|c|c|c|ccc|ccc|ccc}
\hline
\multirow{2}{*}{\begin{tabular}[c]{@{}c@{}}Und.\\ Rat.\end{tabular}}&\multirow{2}{*}{\begin{tabular}[c]{@{}c@{}}Learning\\ Fashion\end{tabular}} & \multirow{2}{*}{Model} & \multirow{2}{*}{\begin{tabular}[c]{@{}c@{}}Training\\ Anatomy\end{tabular}} & \multicolumn{3}{c|}{Knee Reconstruction}                                             & \multicolumn{3}{c|}{Brain  Reconstruction}                                            & \multicolumn{3}{c}{Cardiac  Reconstruction}                                          \\ \cline{5-13} 
                   &                 &                        &                          & \multicolumn{1}{c|}{PSNR}  & \multicolumn{1}{c|}{SSIM}  & $\Delta$ PSNR/SSIM        & \multicolumn{1}{c|}{PSNR}  & \multicolumn{1}{c|}{SSIM}  & $\Delta$ PSNR/SSIM        & \multicolumn{1}{c|}{PSNR}  & \multicolumn{1}{c|}{SSIM}  & $\Delta$ PSNR/SSIM        \\ \hline \hline
\multirow{9}{*}{4x}&\multirow{3}{*}{OAON}                                                       & U-Net                 & K                        & \multicolumn{1}{c|}{34.82} & \multicolumn{1}{c|}{82.09} & \textcolor{red}{0.51/2.24}   & \multicolumn{1}{c|}{34.45} & \multicolumn{1}{c|}{87.24} & \textcolor{blue}{-2.11/-3.92} & \multicolumn{1}{c|}{34.99} & \multicolumn{1}{c|}{85.74} & \textcolor{blue}{-1.56/-3.90} \\  \cline{3-13} 
                                            &                              & U-Net                  & B                        & \multicolumn{1}{c|}{31.12} & \multicolumn{1}{c|}{70.22} & \textcolor{blue}{-3.19/-9.63} & \multicolumn{1}{c|}{36.50} & \multicolumn{1}{c|}{90.83} & \textcolor{blue}{-0.06/-0.33}   & \multicolumn{1}{c|}{33.57} & \multicolumn{1}{c|}{83.85} & \textcolor{blue}{-2.98/-5.79} \\ \cline{3-13} 
                                                    &                        & U-Net                  & C                        & \multicolumn{1}{c|}{33.32} & \multicolumn{1}{c|}{75.64} & \textcolor{blue}{-0.99/-4.21} & \multicolumn{1}{c|}{35.54} & \multicolumn{1}{c|}{89.36} & \textcolor{blue}{-1.02/-1.80} & \multicolumn{1}{c|}{37.17} & \multicolumn{1}{c|}{90.74} & \textcolor{red}{0.62/1.10}   \\ \cline{2-13} 
&MAON                                                       & U-Net & K+B+C               & \multicolumn{1}{c|}{34.31} & \multicolumn{1}{c|}{79.85} & \multicolumn{1}{c|}{0/0}  & \multicolumn{1}{c|}{36.56} & \multicolumn{1}{c|}{91.16} & \multicolumn{1}{c|}{0/0}  & \multicolumn{1}{c|}{36.55} & \multicolumn{1}{c|}{89.64} & \multicolumn{1}{c}{0/0}  \\ \cline{2-13} 
&\multirow{4}{*}{MAPN}                                                       & U-Net+PN1 \cite{liu2021universal}                   & K+B+C                    & \multicolumn{1}{c|}{34.86} & \multicolumn{1}{c|}{81.91} & \textcolor{red}{0.55/2.06}   & \multicolumn{1}{c|}{37.00} & \multicolumn{1}{c|}{91.50} & \textcolor{red}{0.44/0.34}   & \multicolumn{1}{c|}{37.18} & \multicolumn{1}{c|}{90.67} & \textcolor{red}{0.63/1.03}  \\ \cline{3-13} 
                                          &                                  & U-Net+PN2                    & K+B+C                    & \multicolumn{1}{c|}{35.30} & \multicolumn{1}{c|}{\textbf{83.18}} & \textcolor{red}{0.99}/\textcolor{red}{\textbf{3.33}}  & \multicolumn{1}{c|}{37.59} & \multicolumn{1}{c|}{\textbf{92.65}} & \textcolor{red}{1.03/\textbf{1.49}} & \multicolumn{1}{c|}{37.29} & \multicolumn{1}{c|}{90.83} & \textcolor{red}{0.74/1.19}  \\ \cline{3-13} 
                                        &                                    & U-Net+PN3                    & K+B+C                    & \multicolumn{1}{c|}{35.22} & \multicolumn{1}{c|}{82.77} & \textcolor{red}{0.91/2.92}   & \multicolumn{1}{c|}{37.37} & \multicolumn{1}{c|}{91.93} & \textcolor{red}{0.81/0.77}   & \multicolumn{1}{c|}{37.28} & \multicolumn{1}{c|}{90.83} & \textcolor{red}{0.73/1.19}   \\ \cline{3-13} 
                                        &                                    & U-Net+PN4                    & K+B+C                    & \multicolumn{1}{c|}{\textbf{35.36}} & \multicolumn{1}{c|}{83.15} & \textcolor{red}{\textbf{1.05}/3.30}   & \multicolumn{1}{c|}{\textbf{37.66}} & \multicolumn{1}{c|}{92.10} & \textcolor{red}{\textbf{1.10}/0.94}   & \multicolumn{1}{c|}{\textbf{37.30}} & \multicolumn{1}{c|}{\textbf{90.86}} & \textcolor{red}{\textbf{0.75}/\textbf{1.22}}   \\ \hline \hline

\multirow{9}{*}{6x}&\multirow{3}{*}{OAON}                                                       & U-Net                 & K                        & \multicolumn{1}{c|}{33.58} & \multicolumn{1}{c|}{77.65} & \textcolor{red}{0.27/1.32}   & \multicolumn{1}{c|}{32.92} & \multicolumn{1}{c|}{84.21} & \textcolor{blue}{-1.25/-3.34} & \multicolumn{1}{c|}{33.70} & \multicolumn{1}{c|}{83.24} & \textcolor{blue}{-0.89/-2.96} \\  \cline{3-13} 
                                            &                              & U-Net                  & B                        & \multicolumn{1}{c|}{30.70} & \multicolumn{1}{c|}{68.15} & \textcolor{blue}{-2.61/-8.18} & \multicolumn{1}{c|}{34.29} & \multicolumn{1}{c|}{87.83} & \textcolor{red}{0.12/0.28}   & \multicolumn{1}{c|}{32.67} & \multicolumn{1}{c|}{81.76} & \textcolor{blue}{-1.92/-4.44} \\ \cline{3-13} 
                                                    &                        & U-Net                  & C                        & \multicolumn{1}{c|}{32.90} & \multicolumn{1}{c|}{74.02} & \textcolor{blue}{-0.41/-2.31} & \multicolumn{1}{c|}{34.04} & \multicolumn{1}{c|}{87.11} & \textcolor{blue}{-0.13/-0.44} & \multicolumn{1}{c|}{35.46} & \multicolumn{1}{c|}{88.11} & \textcolor{red}{0.87/1.91}   \\ \cline{2-13}
  
&MAON                                                       & U-Net & K+B+C               & \multicolumn{1}{c|}{33.31} & \multicolumn{1}{c|}{76.33} & \multicolumn{1}{c|}{0/0}  & \multicolumn{1}{c|}{34.17} & \multicolumn{1}{c|}{87.55} & \multicolumn{1}{c|}{0/0}  & \multicolumn{1}{c|}{34.59} & \multicolumn{1}{c|}{86.20} & \multicolumn{1}{c}{0/0}  \\ \cline{2-13} 

&\multirow{4}{*}{MAPN}                                                       & U-Net+PN1 \cite{liu2021universal}                    & K+B+C                    & \multicolumn{1}{c|}{33.66} & \multicolumn{1}{c|}{77.76} & \textcolor{red}{0.35/1.43}   & \multicolumn{1}{c|}{34.78} & \multicolumn{1}{c|}{88.66} & \textcolor{red}{0.61/1.11}   & \multicolumn{1}{c|}{35.25} & \multicolumn{1}{c|}{87.71} & \textcolor{red}{0.66/1.51}  \\ \cline{3-13} 
                                          &                                  & U-Net+PN2                    & K+B+C                    & \multicolumn{1}{c|}{\textbf{34.25}} & \multicolumn{1}{c|}{\textbf{79.30}} & \textbf{\textcolor{red}{0.94/2.97}}  & \multicolumn{1}{c|}{\textbf{35.86}} & \multicolumn{1}{c|}{\textbf{90.44}} & \textbf{\textcolor{red}{1.69/2.89}} & \multicolumn{1}{c|}{\textbf{35.80}} & \multicolumn{1}{c|}{\textbf{88.58}} & \textbf{\textcolor{red}{1.21/2.38}}  \\ \cline{3-13} 
                                        &                                    & U-Net+PN3                    & K+B+C                    & \multicolumn{1}{c|}{33.93} & \multicolumn{1}{c|}{78.47} & \textcolor{red}{0.62/2.14}   & \multicolumn{1}{c|}{35.08} & \multicolumn{1}{c|}{89.17} & \textcolor{red}{0.91/1.62}   & \multicolumn{1}{c|}{35.53} & \multicolumn{1}{c|}{88.21} & \textcolor{red}{0.94/2.01}   \\ \cline{3-13} 
                                        &                                    & U-Net+PN4                    & K+B+C                    & \multicolumn{1}{c|}{34.12} & \multicolumn{1}{c|}{79.01} & \textcolor{red}{0.81/2.68}   & \multicolumn{1}{c|}{35.66} & \multicolumn{1}{c|}{{89.91}} & \textcolor{red}{1.49/2.36}   & \multicolumn{1}{c|}{35.64} & \multicolumn{1}{c|}{88.40} & \textcolor{red}{1.05/2.20}   \\
\hline
\end{tabular}}
\end{table*}

\subsection{Experiments of DCCNN based Reconstruction}

Here, we report the reconstruction performance of the anatomy parameterization based DCCNN as shown in Fig. \ref{fig_unet} (b). Experiments show that DCCNN's performance is superior to U-Net in MRI reconstruction, which is consistent with previous works \cite{sun2018compressed,zheng2019cascaded,yan2020neural}. Therefore, we perform more ablation experiments with anatomy parameterization based DCCNN.

\subsubsection{Quantitative Results} The quantitative results are summarized in Table. \ref{tab_1} with the same structure as Table. \ref{tab_3}. Similar to the results of U-Net based reconstruction, we can conclude that networks trained with MAON setting are confused by different anatomies and can not reconstruct MRI as better as the corresponding OAON experts. Our proposed MAPN solutions can learn from different anatomy to better reconstruct MRIs compared to the naive MAON solution or the MAON model with more learning parameters (DCCNN+PN4) - will be discussed later in the following ablation study. Most differently, anatomy-specific channel attention learners (PN2) fail to dig out useful information in most cases in DCCNN based reconstruction. We assume that this is due to the different network architectures between DCCNN and U-Net. Since U-Net's encoder will double the size of feature map after pooling while DCCNN's feature map size is kept constant, the anatomy-specific channel attention may have a better chance in capturing useful information and thus generates better results. The optimal reconstruction performance is obtained by anatomy-specific parallel learners (PN4), which also performs robustly in U-Net based reconstruction results.

\begin{table*}
\caption{Quantitative reconstruction results of DCCNN based reconstruction on 5-fold cross validation of different anatomy parameterization strategies with the metric of PSNR(dB) and SSIM(\%). `Und. Rat.' represents Under-sampling Ratio. `K',`B', and `C' are short for `Knee', `Brain' and `Cardiac' at the 3rd column for compactness. `OAON' is short for `One-Anatomy-One-Network'. `MAON' is short for `Multiple-Anatomy-One-Network'. `MAPN' is short for `Multiple-Anatomy-Parameterized-Network'. The best results are addressed in bold.}
\label{tab_1}
\centering
\scalebox{0.9}{
\begin{tabular}{c|c|c|c|ccc|ccc|ccc}
\hline
\multirow{2}{*}{\begin{tabular}[c]{@{}c@{}}Und.\\ Rat.\end{tabular}}&\multirow{2}{*}{\begin{tabular}[c]{@{}c@{}}Learning\\ Fashion\end{tabular}} & \multirow{2}{*}{Model} & \multirow{2}{*}{\begin{tabular}[c]{@{}c@{}}Training\\ Anatomy\end{tabular}} & \multicolumn{3}{c|}{Knee Reconstruction}                                             & \multicolumn{3}{c|}{Brain  Reconstruction}                                            & \multicolumn{3}{c}{Cardiac  Reconstruction}                                          \\ \cline{5-13} 
                   &                 &                        &                          & \multicolumn{1}{c|}{PSNR}  & \multicolumn{1}{c|}{SSIM}  & $\Delta$ PSNR/SSIM        & \multicolumn{1}{c|}{PSNR}  & \multicolumn{1}{c|}{SSIM}  & $\Delta$ PSNR/SSIM        & \multicolumn{1}{c|}{PSNR}  & \multicolumn{1}{c|}{SSIM}  & $\Delta$ PSNR/SSIM        \\ \hline \hline
\multirow{9}{*}{4x}&\multirow{3}{*}{OAON}                                                       & DCCNN                 & K                        & \multicolumn{1}{c|}{36.08} & \multicolumn{1}{c|}{84.76} & \textcolor{red}{0.29/1.05}   & \multicolumn{1}{c|}{36.54} & \multicolumn{1}{c|}{90.26} & \textcolor{blue}{-2.14/-2.68} & \multicolumn{1}{c|}{37.73} & \multicolumn{1}{c|}{90.76} & \textcolor{blue}{-1.62/-2.37} \\  \cline{3-13} 
                                            &                              & DCCNN                  & B                        & \multicolumn{1}{c|}{33.45} & \multicolumn{1}{c|}{76.14} & \textcolor{blue}{-2.34/-7.57} & \multicolumn{1}{c|}{39.19} & \multicolumn{1}{c|}{93.38} & \textcolor{red}{0.51/0.44}   & \multicolumn{1}{c|}{36.99} & \multicolumn{1}{c|}{88.94} & \textcolor{blue}{-2.36/-4.19} \\ \cline{3-13} 
                                                    &                        & DCCNN                  & C                        & \multicolumn{1}{c|}{34.75} & \multicolumn{1}{c|}{80.47} & \textcolor{blue}{-1.04/-3.24} & \multicolumn{1}{c|}{37.74} & \multicolumn{1}{c|}{91.79} & \textcolor{blue}{-0.94/-1.15} & \multicolumn{1}{c|}{40.30} & \multicolumn{1}{c|}{94.25} & \textcolor{red}{0.95/1.12}   \\ \cline{2-13} 
&\multirow{2}{*}{MAON}                                                       & DCCNN & K+B+C               & \multicolumn{1}{c|}{35.79} & \multicolumn{1}{c|}{83.71} & \multicolumn{1}{c|}{0/0}  & \multicolumn{1}{c|}{38.68} & \multicolumn{1}{c|}{92.94} & \multicolumn{1}{c|}{0/0}  & \multicolumn{1}{c|}{39.35} & \multicolumn{1}{c|}{93.13} & \multicolumn{1}{c}{0/0}  \\ \cline{3-13} 
                                         &                                   & DCCNN+PN4 & K+B+C                                                                            & \multicolumn{1}{c|}{35.49} & \multicolumn{1}{c|}{82.91} & \textcolor{blue}{-0.30/-0.80}               & \multicolumn{1}{c|}{38.25} & \multicolumn{1}{c|}{92.61} & \textcolor{blue}{-0.43/-0.33}               & \multicolumn{1}{c|}{38.82} & \multicolumn{1}{c|}{92.41} & \textcolor{blue}{-0.53/-0.72} \\ \cline{2-13} 

&\multirow{4}{*}{MAPN}                                                       & DCCNN+PN1 \cite{liu2021universal}                   & K+B+C                    & \multicolumn{1}{c|}{35.95} & \multicolumn{1}{c|}{84.33} & \textcolor{red}{0.16/0.62}   & \multicolumn{1}{c|}{38.91} & \multicolumn{1}{c|}{93.19} & \textcolor{red}{0.23/0.25}   & \multicolumn{1}{c|}{39.57} & \multicolumn{1}{c|}{93.36} & \textcolor{red}{0.21/0.23}  \\ \cline{3-13} 
                                          &                                  & DCCNN+PN2                    & K+B+C                    & \multicolumn{1}{c|}{35.75} & \multicolumn{1}{c|}{84.14} & \textcolor{blue}{-0.04}/\textcolor{red}{0.43}  & \multicolumn{1}{c|}{38.39} & \multicolumn{1}{c|}{92.82} & \textcolor{blue}{-0.29/-0.12} & \multicolumn{1}{c|}{38.92} & \multicolumn{1}{c|}{92.61} & \textcolor{blue}{-0.43/-0.52}  \\ \cline{3-13} 
                                        &                                    & DCCNN+PN3                    & K+B+C                    & \multicolumn{1}{c|}{36.31} & \multicolumn{1}{c|}{85.35} & \textcolor{red}{0.52/1.64}   & \multicolumn{1}{c|}{39.86} & \multicolumn{1}{c|}{93.92} & \textcolor{red}{1.18/0.98}   & \multicolumn{1}{c|}{40.31} & \multicolumn{1}{c|}{94.25} & \textcolor{red}{0.96/1.12}   \\ \cline{3-13} 
                                        &                                    & DCCNN+PN4                    & K+B+C                    & \multicolumn{1}{c|}{\textbf{36.50}} & \multicolumn{1}{c|}{\textbf{85.82}} & \textbf{\textcolor{red}{0.71/2.11}}   & \multicolumn{1}{c|}{\textbf{40.15}} & \multicolumn{1}{c|}{\textbf{94.32}} & \textbf{\textcolor{red}{1.47/1.38}}   & \multicolumn{1}{c|}{\textbf{40.49}} & \multicolumn{1}{c|}{\textbf{94.45}} & \textbf{\textcolor{red}{1.14/1.32}}   \\ \hline \hline

\multirow{9}{*}{6x}&\multirow{3}{*}{OAON}                                                       & DCCNN                 & K                        & \multicolumn{1}{c|}{34.13} & \multicolumn{1}{c|}{78.77} & \textcolor{red}{0.19/0.78}   & \multicolumn{1}{c|}{33.78} & \multicolumn{1}{c|}{85.47} & \textcolor{blue}{-1.39/-3.06} & \multicolumn{1}{c|}{35.02} & \multicolumn{1}{c|}{86.18} & \textcolor{blue}{-1.17/-2.61} \\  \cline{3-13} 
                                            &                              & DCCNN                  & B                        & \multicolumn{1}{c|}{32.12} & \multicolumn{1}{c|}{71.49} & \textcolor{blue}{-1.82/-6.50} & \multicolumn{1}{c|}{35.64} & \multicolumn{1}{c|}{89.34} & \textcolor{red}{0.47/0.81}   & \multicolumn{1}{c|}{34.78} & \multicolumn{1}{c|}{85.07} & \textcolor{blue}{-1.41/-3.72} \\ \cline{3-13} 
                                                    &                        & DCCNN                  & C                        & \multicolumn{1}{c|}{33.39} & \multicolumn{1}{c|}{75.95} & \textcolor{blue}{-0.55/-2.04} & \multicolumn{1}{c|}{34.72} & \multicolumn{1}{c|}{87.61} & \textcolor{blue}{-0.45/-0.92} & \multicolumn{1}{c|}{36.85} & \multicolumn{1}{c|}{90.05} & \textcolor{red}{0.66/1.26}   \\ \cline{2-13}
  
&MAON                                                       & DCCNN & K+B+C               & \multicolumn{1}{c|}{33.94} & \multicolumn{1}{c|}{77.99} & \multicolumn{1}{c|}{0/0}  & \multicolumn{1}{c|}{35.17} & \multicolumn{1}{c|}{88.53} & \multicolumn{1}{c|}{0/0}  & \multicolumn{1}{c|}{36.19} & \multicolumn{1}{c|}{88.79} & \multicolumn{1}{c}{0/0}  \\ \cline{2-13} 

&\multirow{4}{*}{MAPN}                                                       & DCCNN+PN1 \cite{liu2021universal}                    & K+B+C                    & \multicolumn{1}{c|}{34.05} & \multicolumn{1}{c|}{78.48} & \textcolor{red}{0.11/0.49}   & \multicolumn{1}{c|}{35.45} & \multicolumn{1}{c|}{89.07} & \textcolor{red}{0.28/0.54}   & \multicolumn{1}{c|}{36.48} & \multicolumn{1}{c|}{89.29} & \textcolor{red}{0.29/0.50}  \\ \cline{3-13} 
                                          &                                  & DCCNN+PN2                    & K+B+C                    & \multicolumn{1}{c|}{34.11} & \multicolumn{1}{c|}{78.71} & \textcolor{red}{0.17}/\textcolor{red}{0.72}  & \multicolumn{1}{c|}{35.23} & \multicolumn{1}{c|}{88.83} & \textcolor{red}{0.06/0.30} & \multicolumn{1}{c|}{36.10} & \multicolumn{1}{c|}{88.66} & \textcolor{blue}{-0.09/-0.13}  \\ \cline{3-13} 
                                        &                                    & DCCNN+PN3                    & K+B+C                    & \multicolumn{1}{c|}{34.28} & \multicolumn{1}{c|}{79.24} & \textcolor{red}{0.34/1.25}   & \multicolumn{1}{c|}{36.17} & \multicolumn{1}{c|}{90.21} & \textcolor{red}{1.00/1.68}   & \multicolumn{1}{c|}{36.87} & \multicolumn{1}{c|}{90.06} & \textcolor{red}{0.68/1.27}   \\ \cline{3-13} 
                                        &                                    & DCCNN+PN4                    & K+B+C                    & \multicolumn{1}{c|}{\textbf{34.47}} & \multicolumn{1}{c|}{\textbf{79.75}} & \textbf{\textcolor{red}{0.53/1.76}}   & \multicolumn{1}{c|}{\textbf{36.51}} & \multicolumn{1}{c|}{\textbf{90.77}} & \textbf{\textcolor{red}{1.34/2.24}}   & \multicolumn{1}{c|}{\textbf{37.07}} & \multicolumn{1}{c|}{\textbf{90.39}} & \textbf{\textcolor{red}{0.88/1.60}}   \\
\hline
\end{tabular}}
\end{table*}

\begin{figure*}
\centering
\includegraphics[width=\textwidth]{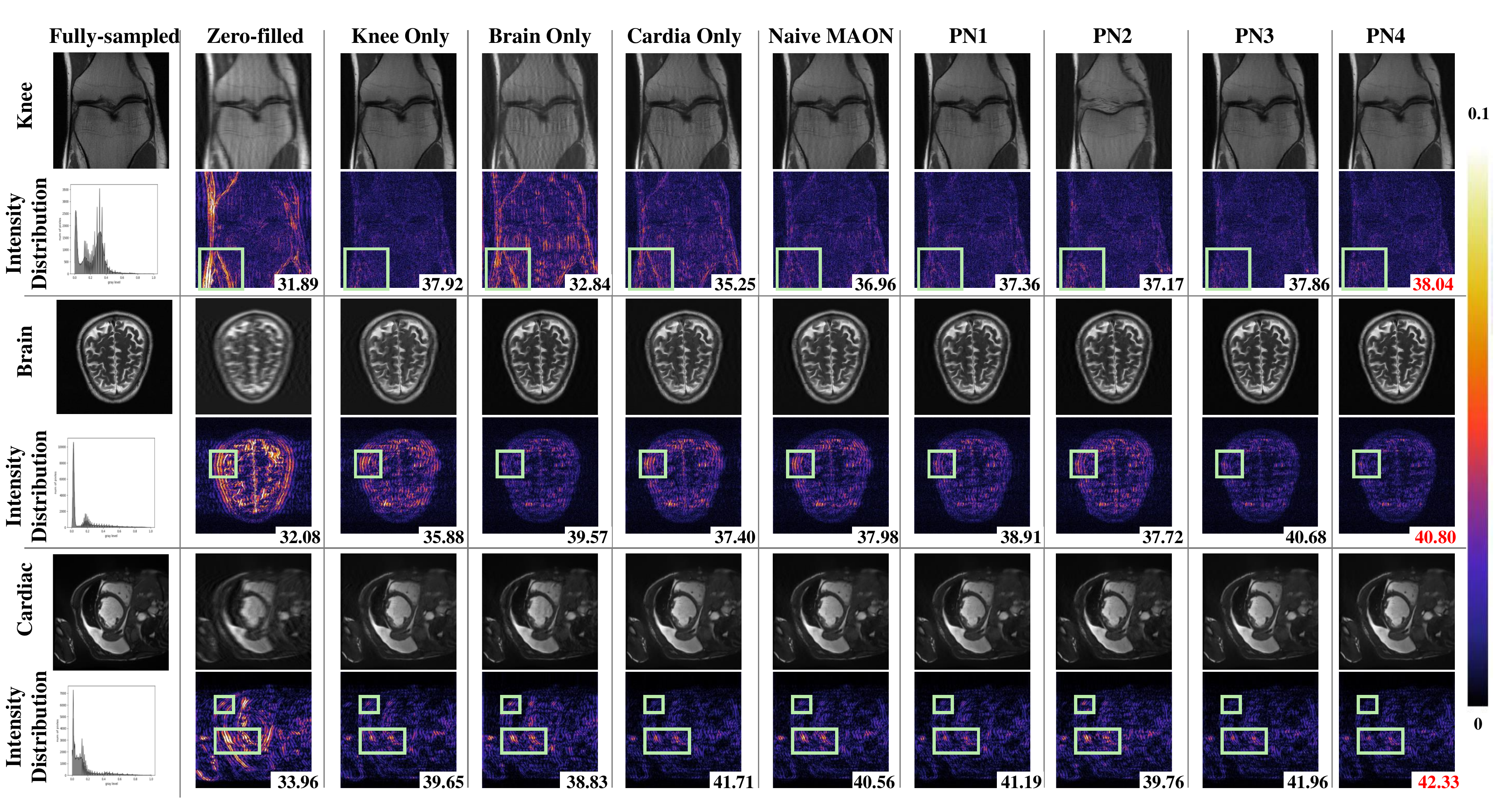}
\caption{Qualitative comparison of different anatomy parameterization DCCNN reconstruction settings with $4\times$ under-sampling ratio. Original MRIs are normalized to [0, 1] for visualization and their intensity distributions are shown in the first column. From the second column, reconstructed image (top) and its absolute error map (bottom) are shown. PSNR (dB) is listed at the right bottom of each error map with the highest one highlighted in red. Values $>0.1$ in error maps are clipped to 0.1 for better visual comparison. Green boxes address where significant performance difference can be observed.}
\label{fig_comparison_bn}
\end{figure*}

\subsubsection{Qualitative Results} We also visually compare the reconstruction performance of all settings in DCCNN with $4\times$ ratio in Fig. \ref{fig_comparison_bn}. We can also find that the visual quality of final reconstructed MRIs are improved with our strategies.

\begin{figure*}
  \centering
  \centerline{\includegraphics[width=\textwidth]{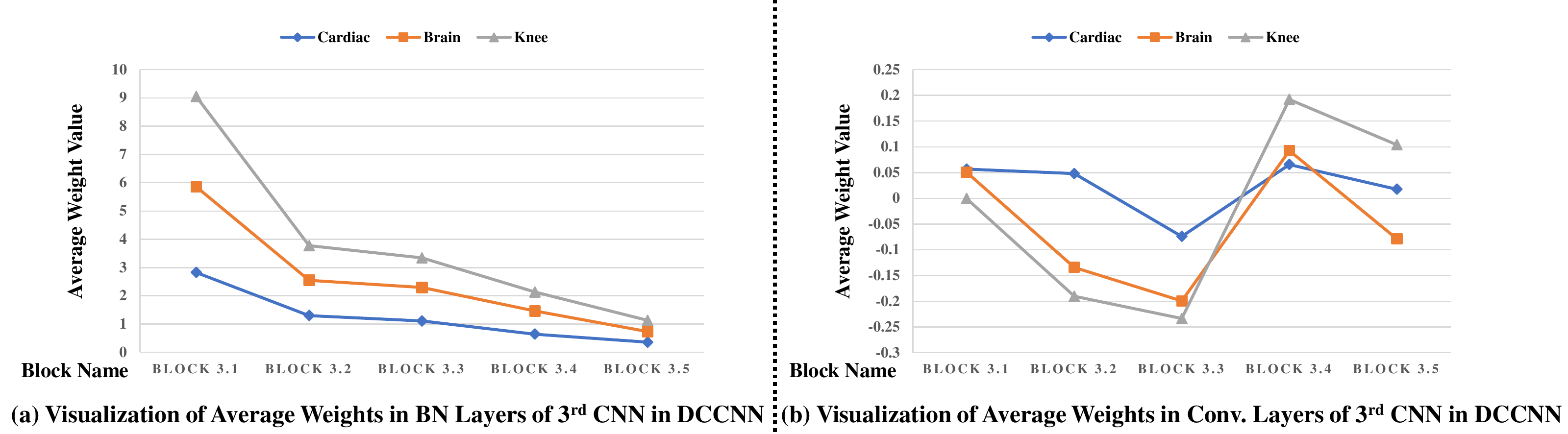}}
  \caption{Visualization of learned parameters in the 3rd CNN of DCCNN+PN4 model. The average weight values in BN of 5 blocks are shown in (a), and the average weight values of $1\times1$ convolutional layers are reported in (b). The weights in additive learning modules are anatomy-sensitive. The weights of BN can re-scale different distributions for different anatomies with simple value shifting. The weights of anatomy-specific $1\times1$ convolutional layers vary more significantly between each other and help to improve the final results.}
  \label{fig_visp}
\end{figure*}

\begin{table}
\caption{Comparison of learning parameter number between different anatomy parameterization settings with DCCNN. Note that the parameter sum of MAON+DCCNN+PN4 is not exactly equal to MAPN+DCCNN+PN4 due to anatomy-shared BN settings.}
\label{tab_2}
\centering
\scalebox{0.8}{
\begin{tabular}{c|c|c|c|c}
\hline
\begin{tabular}[c]{@{}c@{}}Learning\\ Fashion\end{tabular} & Model     & \begin{tabular}[c]{@{}c@{}}Anatomy-shared \\ Parameters (K)\end{tabular} & \begin{tabular}[c]{@{}c@{}}Anatomy-specific \\ Parameters (K) \end{tabular} & \begin{tabular}[c]{@{}c@{}}Three Antomy\\ Param. Sum (K)\end{tabular} \\ \hline
OAON                                                       & DCCNN     & 0.00                                                                 & 569.64                                                                & 1708.92                                                           \\ \hline
\multirow{2}{*}{MAON}                                      & DCCNN     & 569.64                                                              & 0.00                                                                   & 569.64                                                           \\ \cline{2-5} 
                                                           & DCCNN+PN4 & 757.80                                                              & 0.00                                                                   & 757.80                                                           \\ \hline
\multirow{4}{*}{MAPN}                                      & DCCNN+PN1 \cite{liu2021universal}  & 564.48                                                              & 5.16                                                                  & 579.96                                                           \\ \cline{2-5} 
                                                           & DCCNN+PN2 & 564.48                                                              & 87.10                                                                 & 825.78                                                           \\ \cline{2-5} 
                                                           & DCCNN+PN3 & 564.48                                                              & 89.68                                                                 & 833.52                                                           \\ \cline{2-5} 
                                                           & DCCNN+PN4 & 564.48                                                              & 67.88                                                                 & 768.12                                                           \\ \hline
\end{tabular}}
\end{table}

\subsubsection{Comparison on the Increment of Learning Parameters} There is a remaining question that whether the improvement is brought by the increment of learning parameters. We re-draw the MAPN+PN4 in Fig.\ref{fig_framework} (d-i) as MAON+PN4 (d-ii), where 3 paralleled $1 \times 1$ convolutional layers are added, and we trained the DCCNN+PN4 with the MAON setting with the quantitative results reported in Table. \ref{tab_1} with $4\times$ ratio. Here, we do not use 3 paralleled BN layers because such setting seems to be non-sense. It can be concluded that MAON+DCCNN+PN4 results in worse reconstruction than MAPN+DCCNN+PN4, suggesting that the performance gain achieved by MAPN+PN4 is not due to more parameters nor the parallel convolution with different receptivity fields. The learning parameter numbers of all the involved anatomy parameterization settings are summarized in Table. \ref{tab_2}. For a fair comparison, OAON setting needs 3 expert models with $3 \times 569.64$K = $1708.92$K parameters in total to outperform the MAON baseline with $569.64$K parameters for three anatomy reconstruction tasks. While our MAPN+PN4 models contains $768.12$K parameters, which can save $>50\%$ parameters with better results than OAON. The MAON+PN4 model has $757.80$K parameters, which are not exactly equal to MAPN+PN4 without 3 paralleled BN in every convolution blocks. 

\subsubsection{Comparison on the Inference Time} We also evaluate the inference time for different methods, and PN1-4 show little difference from the original DCCNN due to PN1-4 are parameter-efficient. It takes $<0.6s$ to reconstruct per slice for all the methods with an NVIDIA GeForce RTX 3090 GPU and Pytorch backend.

\subsubsection{Visualization of the Unique Information Captured by Anatomy-specific Learners} Here, we try to discuss what the simple anatomy-specific learners (BN and $1\times1$ convolution layers) capture for better reconstruction. However, since MRI reconstruction is a low-level signal reconstruction task, to visualize and explain the feature maps inside the network remains an unsolved problem itself. So, we visualize the final learned parameters in the anatomy-specific learners.

%\subsubsection{Visualization of Learned Parameters}
For simplicity, we focus on the parameters in DCCNN+PN4, which achieves the best performance in our experiments. Without loss of generalization ability, we visualize the learned parameters of the 3rd CNN by the following strategy: we calculate and present the average value of weights in anatomy-specific BN and $1 \times 1$ convolutional operations across 5 blocks, and show them in Fig.\ref{fig_visp} (a) and (b). 

We can find that the weights in additive learning modules are anatomy-sensitive. The weights of BN can re-scale different distributions for different anatomies with simple value shifting. The weights of anatomy-specific $1\times1$ convolutional layers vary more significantly between each other compared to anatomy-specific BN layers, indicating that paralleled $1\times1$ learners do learn something more unique than BN layers. As a result, the integrated paralleled learners can then effectively guide the reconstruction for their targeted anatomy and finally contribute to the overall improvement. 

%\subsubsection{Visualization of Learned Feature Maps} Although the feature maps may be difficult to explain, we also try to visualize of learned feature maps with the following strategy: we focus on the 3rd layer in the 3rd CNN, termed as the middle level, by calculating the maximum value from 64 feature maps before/after every operation in the PN4 module, shown in Fig.\ref{fig_visp} (c). Given one specific anatomy instance, the maximum feature value maps are normalized with their own range and augmented by the same max-min window for displaying visual quality. We can see that the anatomy-shared stream, Fig.\ref{fig_visp} (a)-(II), captures shared de-aliasing patterns with significant wave-like patterns while the anatomy-specific stream, Fig.\ref{fig_visp} (a)-(III), enhances unique anatomy structures without wave-like patterns.

\section{Conclusion and Discussion}
In this work, we focus on an under-explored problem that how to make different anatomy to collaborate with each other in one network for better MRI reconstruction results. We reformulate the common deep MRI reconstruction network with anatomy-shared and anatomy-specific learners and build a novel Multiple-Anatomy-Parameterized-Network framework. Under our proposed framework, we explore four possible implementations of anatomy-specific learners and evaluate their performance on three anatomies with U-Net or DCCNN based reconstruction networks. Experimental results prove that our framework can help to boost the final performance in multiple anatomy reconstruction tasks in general for different networks. It is surprising that anatomy-specific learners built with simple BN layers and $1 \times 1$ convolution can capture useful information, which is evident in not only the reconstruction results but also the visualization of learned parameters inside these learners. However, under our proposed MAPN framework, how to design the optimal anatomy-specific learners is still an open problem, and we believe this work sheds the light in a new direction towards more robust MRI reconstruction for multiple anatomy scene.

Our work is closely related to previous works \cite{yan2020neural} and \cite{liu2021universal}. In \cite{yan2020neural}, researchers found that the optimal network architectures obtained via neural architecture searching are different for brain and knee reconstruction tasks. However, it is difficult and impractical to design or search an anatomy-specific architecture for every anatomy. In this work, we solve this problem by disentangling the whole network into anatomy-share and anatomy-specific parts and using the flexible anatomy-specific learners to model unique knowledge instead of changing the entire architecture. We also reveal that the same anatomy-specific learners, especially PN2, perform differently when they are integrated into different reconstruction network architectures (U-Net and DCCNN). Based on this observation, one potential future extension of our work is to search the architectures of the flexible anatomy-specific learners via neural architecture searching. Another important work towards a universal MRI reconstruction is \cite{liu2021universal}, where the researchers also used the anatomy-specific normalization settings (PN1). However, they mainly focused on using knowledge distillation in DCCNN to enhance the final performance, where well pre-trained expert networks are required before training the universal model. In contrast, our framework does not require any pre-trained expert models, and we further enhance anatomy-specific normalization with more advanced designed learners. Experiments show that our settings can robustly handle different networks.

Despite the good results, current experiments are still early attempts of our proposed framework, and there exist some limitations. First, current experiments are only conducted in single-coil MRI reconstruction scenario for demonstration purpose. As multi-coil MRI reconstruction \cite{sriram2020grappanet,jun2021joint} is more popular in clinical practice, extending this work to multi-coil reconstruction will be an important future work. Since we have evaluated our framework with DCCNN \cite{schlemper2017deep}, it is easy to extend DCCNN \cite{aggarwal2018modl} to MoDL \cite{aggarwal2018modl} framework by modifying the network slightly and using conjugate gradient method to solve the Eq. \ref{equ:m} for multi-coil MRIs. Another limitation lies in the objective loss function. In this work, we only adopted L1 loss to supervise the learning process of deep networks for a fair and initial comparison without considering the real distribution of the noise. It will be an important future work to introduce more robust objective loss functions \cite{yang2017dagan,cheng2021learning,zhou2022accelerating} to supervise the designed learners for better modeling the reconstruction process in practice.

%\section{Conclusion}
%In this work, we aim to enhance the anatomy generalization ability of deep undersampled MRI reconstruction network. We propose a novel Multiple-Anatomy-Parameterized-Network framework by re-architecturing the deep MRI reconstruction network with anatomy-shared/specific learners. Four compact anatomy-specific learners are designed and explored to learn exclusive anatomy knowledge. Experiments on three anatomy MRI datasets show that these learners can help different anatomies to collaborate with rather than harm each other in different reconstruction networks for overall reconstruction performance improvement, proving the potential of our solution. 

\section*{Acknowledgment}
The authors would like to thank the fastMRI \cite{fastMRI} project for making their codes and datasets accessible online.

\bibliographystyle{IEEEtran.bst}
\input{main.bbl}

\end{document}

%% file: main.bbl

%% file: main.bbl
\begin{thebibliography}{10}
\providecommand{\url}[1]{#1}
\csname url@samestyle\endcsname
\providecommand{\newblock}{\relax}
\providecommand{\bibinfo}[2]{#2}
\providecommand{\BIBentrySTDinterwordspacing}{\spaceskip=0pt\relax}
\providecommand{\BIBentryALTinterwordstretchfactor}{4}
\providecommand{\BIBentryALTinterwordspacing}{\spaceskip=\fontdimen2\font plus
\BIBentryALTinterwordstretchfactor\fontdimen3\font minus
  \fontdimen4\font\relax}
\providecommand{\BIBforeignlanguage}[2]{{%
\expandafter\ifx\csname l@#1\endcsname\relax
\typeout{** WARNING: IEEEtran.bst: No hyphenation pattern has been}%
\typeout{** loaded for the language `#1'. Using the pattern for}%
\typeout{** the default language instead.}%
\else
\language=\csname l@#1\endcsname
\fi
#2}}
\providecommand{\BIBdecl}{\relax}
\BIBdecl

\bibitem{donoho2006compressed}
D.~L. Donoho, ``Compressed sensing,'' \emph{IEEE Transactions on Information
  Theory}, vol.~52, no.~4, pp. 1289--1306, 2006.

\bibitem{liang2009accelerating}
D.~Liang, B.~Liu, J.~Wang, and L.~Ying, ``Accelerating sense using compressed
  sensing,'' \emph{Magnetic Resonance in Medicine}, vol.~62, no.~6, pp.
  1574--1584, 2009.

\bibitem{huang2011efficient}
J.~Huang, S.~Zhang, and D.~Metaxas, ``Efficient mr image reconstruction for
  compressed mr imaging,'' \emph{Medical Image Analysis}, vol.~15, no.~5, pp.
  670--679, 2011.

\bibitem{liang2020deep}
D.~Liang, J.~Cheng, Z.~Ke, and L.~Ying, ``Deep magnetic resonance image
  reconstruction: Inverse problems meet neural networks,'' \emph{IEEE Signal
  Processing Magazine}, vol.~37, no.~1, pp. 141--151, 2020.

\bibitem{knoll2020deep}
F.~Knoll, K.~Hammernik, C.~Zhang, S.~Moeller, T.~Pock, D.~K. Sodickson, and
  M.~Akcakaya, ``Deep-learning methods for parallel magnetic resonance imaging
  reconstruction: A survey of the current approaches, trends, and issues,''
  \emph{IEEE Signal Processing Magazine}, vol.~37, no.~1, pp. 128--140, 2020.

\bibitem{wang2016accelerating}
S.~Wang, Z.~Su, L.~Ying, X.~Peng, S.~Zhu, F.~Liang, D.~Feng, and D.~Liang,
  ``Accelerating magnetic resonance imaging via deep learning,'' in \emph{2016
  IEEE 13th International Symposium on Biomedical Imaging}.\hskip 1em plus
  0.5em minus 0.4em\relax IEEE, 2016, pp. 514--517.

\bibitem{schlemper2017deep}
J.~Schlemper, J.~Caballero, J.~V. Hajnal, A.~N. Price, and D.~Rueckert, ``A
  deep cascade of convolutional neural networks for dynamic mr image
  reconstruction,'' \emph{IEEE Transactions on Medical Imaging}, vol.~37,
  no.~2, pp. 491--503, 2017.

\bibitem{qin2018convolutional}
C.~Qin, J.~Schlemper, J.~Caballero, A.~N. Price, J.~V. Hajnal, and D.~Rueckert,
  ``Convolutional recurrent neural networks for dynamic mr image
  reconstruction,'' \emph{IEEE Transactions on Medical Imaging}, vol.~38,
  no.~1, pp. 280--290, 2018.

\bibitem{hammernik2018learning}
K.~Hammernik, T.~Klatzer, E.~Kobler, M.~P. Recht, D.~K. Sodickson, T.~Pock, and
  F.~Knoll, ``Learning a variational network for reconstruction of accelerated
  mri data,'' \emph{Magnetic Resonance in Medicine}, vol.~79, no.~6, pp.
  3055--3071, 2018.

\bibitem{aggarwal2018modl}
H.~K. Aggarwal, M.~P. Mani, and M.~Jacob, ``Modl: Model-based deep learning
  architecture for inverse problems,'' \emph{IEEE Transactions on Medical
  Imaging}, vol.~38, no.~2, pp. 394--405, 2018.

\bibitem{lee2018deep}
D.~Lee, J.~Yoo, S.~Tak, and J.~C. Ye, ``Deep residual learning for accelerated
  mri using magnitude and phase networks,'' \emph{IEEE Transactions on
  Biomedical Engineering}, vol.~65, no.~9, pp. 1985--1995, 2018.

\bibitem{sun2018compressed}
L.~Sun, Z.~Fan, Y.~Huang, X.~Ding, and J.~Paisley, ``Compressed sensing mri
  using a recursive dilated network,'' in \emph{Thirty-Second AAAI Conference
  on Artificial Intelligence}, 2018.

\bibitem{zheng2019cascaded}
H.~Zheng, F.~Fang, and G.~Zhang, ``Cascaded dilated dense network with two-step
  data consistency for mri reconstruction,'' in \emph{Advances in Neural
  Information Processing Systems}, 2019.

\bibitem{yan2020neural}
J.~Yan, S.~Chen, Y.~Zhang, and X.~Li, ``Neural architecture search for
  compressed sensing magnetic resonance image reconstruction,''
  \emph{Computerized Medical Imaging and Graphics}, vol.~85, p. 101784, 2020.

\bibitem{sriram2020end}
A.~Sriram, J.~Zbontar, T.~Murrell, A.~Defazio, C.~L. Zitnick, N.~Yakubova,
  F.~Knoll, and P.~Johnson, ``End-to-end variational networks for accelerated
  mri reconstruction,'' in \emph{International Conference on Medical Image
  Computing and Computer Assisted Intervention}.\hskip 1em plus 0.5em minus
  0.4em\relax Springer, 2020, pp. 64--73.

\bibitem{jun2021joint}
Y.~Jun, H.~Shin, T.~Eo, and D.~Hwang, ``Joint deep model-based mr image and
  coil sensitivity reconstruction network (joint-icnet) for fast mri,'' in
  \emph{Proceedings of the IEEE/CVF Conference on Computer Vision and Pattern
  Recognition}, 2021, pp. 5270--5279.

\bibitem{liu2021universal}
X.~Liu, J.~Wang, F.~Liu, and S.~K. Zhou, ``Universal undersampled mri
  reconstruction,'' in \emph{International Conference on Medical Image
  Computing and Computer Assisted Intervention}.\hskip 1em plus 0.5em minus
  0.4em\relax Springer, 2021, pp. 211--221.

\bibitem{rebuffi17}
S.-A. Rebuffi, H.~Bilen, and A.~Vedaldi, ``Learning multiple visual domains
  with residual adapters,'' in \emph{Advances in Neural Information Processing
  Systems}, 2017.

\bibitem{rebuffi18}
------, ``Efficient parametrization of multi-domain deep neural networks,'' in
  \emph{Proceedings of the IEEE Conference on Computer Vision and Pattern
  Recognition}, 2018, pp. 8119--8127.

\bibitem{fastMRI}
J.~Zbontar, F.~Knoll, A.~Sriram, M.~J. Muckley, M.~Bruno, A.~Defazio,
  M.~Parente, K.~Geras, J.~Katsnelson, H.~Chandarana, Z.~Zhang, M.~Drozdzal,
  A.~Romero, M.~G. Rabbat, P.~Vincent, J.~Pinkerton, D.~Wang, N.~Yakubova,
  E.~Owens, C.~L. Zitnick, M.~P. Recht, D.~K. Sodickson, and Y.~W. Lui,
  ``fastmri: An open dataset and benchmarks for accelerated mri,''
  \emph{ArXiv}, vol. abs/1811.08839, 2018.

\bibitem{zhu2018image}
B.~Zhu, J.~Z. Liu, S.~F. Cauley, B.~R. Rosen, and M.~S. Rosen, ``Image
  reconstruction by domain-transform manifold learning,'' \emph{Nature}, vol.
  555, no. 7697, pp. 487--492, 2018.

\bibitem{yang2017dagan}
G.~Yang, S.~Yu, H.~Dong, G.~Slabaugh, P.~L. Dragotti, X.~Ye, F.~Liu,
  S.~Arridge, J.~Keegan, Y.~Guo \emph{et~al.}, ``Dagan: deep de-aliasing
  generative adversarial networks for fast compressed sensing mri
  reconstruction,'' \emph{IEEE Transactions on Medical Imaging}, vol.~37,
  no.~6, pp. 1310--1321, 2017.

\bibitem{akccakaya2019scan}
M.~Ak{\c{c}}akaya, S.~Moeller, S.~Weing{\"a}rtner, and K.~U{\u{g}}urbil,
  ``Scan-specific robust artificial-neural-networks for k-space interpolation
  (raki) reconstruction: Database-free deep learning for fast imaging,''
  \emph{Magnetic Resonance in Medicine}, vol.~81, no.~1, pp. 439--453, 2019.

\bibitem{zhang2021dual}
Y.~Zhang, J.~Lyu, and X.~Bi, ``A dual-task dual-domain model for blind mri
  reconstruction,'' \emph{Computerized Medical Imaging and Graphics}, vol.~89,
  p. 101862, 2021.

\bibitem{ARSHAD202196}
M.~Arshad, M.~Qureshi, O.~Inam, and H.~Omer, ``Transfer learning in deep neural
  network based under-sampled mr image reconstruction,'' \emph{Magnetic
  Resonance Imaging}, vol.~76, pp. 96--107, 2021.

\bibitem{LV2021104504}
J.~Lv, G.~Li, X.~Tong, W.~Chen, J.~Huang, C.~Wang, and G.~Yang, ``Transfer
  learning enhanced generative adversarial networks for multi-channel mri
  reconstruction,'' \emph{Computers in Biology and Medicine}, vol. 134, p.
  104504, 2021.

\bibitem{dar2020transfer}
S.~U.~H. Dar, M.~{\"O}zbey, A.~B. {\c{C}}atl{\i}, and T.~{\c{C}}ukur, ``A
  transfer-learning approach for accelerated mri using deep neural networks,''
  \emph{Magnetic Resonance in Medicine}, vol.~84, no.~2, pp. 663--685, 2020.

\bibitem{guo2021multi}
P.~Guo, P.~Wang, J.~Zhou, S.~Jiang, and V.~M. Patel, ``Multi-institutional
  collaborations for improving deep learning-based magnetic resonance image
  reconstruction using federated learning,'' in \emph{Proceedings of the
  IEEE/CVF Conference on Computer Vision and Pattern Recognition}, 2021, pp.
  2423--2432.

\bibitem{bilen2017universal}
H.~Bilen and A.~Vedaldi, ``Universal representations: The missing link between
  faces, text, planktons, and cat breeds,'' \emph{arXiv preprint
  arXiv:1701.07275}, 2017.

\bibitem{mallya2018piggyback}
A.~Mallya, D.~Davis, and S.~Lazebnik, ``Piggyback: Adapting a single network to
  multiple tasks by learning to mask weights,'' in \emph{Proceedings of the
  European Conference on Computer Vision}, 2018, pp. 67--82.

\bibitem{ronneberger2015u}
O.~Ronneberger, P.~Fischer, and T.~Brox, ``U-net: Convolutional networks for
  biomedical image segmentation,'' in \emph{International Conference on Medical
  image computing and computer Assisted intervention}.\hskip 1em plus 0.5em
  minus 0.4em\relax Springer, 2015, pp. 234--241.

\bibitem{ioffe2015batch}
S.~Ioffe and C.~Szegedy, ``Batch normalization: Accelerating deep network
  training by reducing internal covariate shift,'' in \emph{International
  Conference on Machine Learning}.\hskip 1em plus 0.5em minus 0.4em\relax PMLR,
  2015, pp. 448--456.

\bibitem{hu2018squeeze}
J.~Hu, L.~Shen, and G.~Sun, ``Squeeze-and-excitation networks,'' in
  \emph{Proceedings of the IEEE Conference on Computer Vision and Pattern
  Recognition}, 2018, pp. 7132--7141.

\bibitem{bernard2018deep}
O.~Bernard, A.~Lalande, C.~Zotti, F.~Cervenansky, X.~Yang, P.-A. Heng,
  I.~Cetin, K.~Lekadir, O.~Camara, M.~A.~G. Ballester \emph{et~al.}, ``Deep
  learning techniques for automatic mri cardiac multi-structures segmentation
  and diagnosis: Is the problem solved?'' \emph{IEEE Transactions on Medical
  Imaging}, vol.~37, no.~11, pp. 2514--2525, 2018.

\bibitem{hore2010image}
A.~Hore and D.~Ziou, ``Image quality metrics: Psnr vs. ssim,'' in \emph{2010
  20th International Conference on Pattern Recognition}.\hskip 1em plus 0.5em
  minus 0.4em\relax IEEE, 2010, pp. 2366--2369.

\bibitem{sriram2020grappanet}
A.~Sriram, J.~Zbontar, T.~Murrell, C.~L. Zitnick, A.~Defazio, and D.~K.
  Sodickson, ``Grappanet: Combining parallel imaging with deep learning for
  multi-coil mri reconstruction,'' in \emph{Proceedings of the IEEE/CVF
  Conference on Computer Vision and Pattern Recognition}, 2020, pp.
  14\,315--14\,322.

\bibitem{cheng2021learning}
J.~Cheng, Z.-X. Cui, W.~Huang, Z.~Ke, L.~Ying, H.~Wang, Y.~Zhu, and D.~Liang,
  ``Learning data consistency and its application to dynamic mr imaging,''
  \emph{IEEE Transactions on Medical Imaging}, vol.~40, no.~11, pp. 3140--3153,
  2021.

\bibitem{zhou2022accelerating}
Y.~Zhou, H.~Wang, Y.~Liu, D.~Liang, and L.~Ying, ``Accelerating mr parameter
  mapping using nonlinear compressive manifold learning and regularized
  pre-imaging,'' \emph{IEEE Transactions on Biomedical Engineering}, 2022.

\end{thebibliography}
